# Nested interaction networks represent a missing link in the study of behavioural and community ecology

The EcoEvoInteract Scientific Network

*Authors:*

Montiglio, P.O.[1,*], Gotanda, K.M.[2], Kratochwil, C.F.[3,4], Laskowski, K.L.[5], Nadell, C.D.[6], Farine, D.R.[7,8,9,*]

[1]Department of Biology and Redpath Museum, McGill University

[2]Department of Zoology, University of Cambridge

[3]Chair in Zoology and Evolutionary Biology, Department of Biology, University of Konstanz

[4]Zukunftskolleg, University of Konstanz, Konstanz, Germany

[5]Department of Biology & Ecology of Fishes, Leibniz-Institute of Freshwater Ecology & Inland Fisheries

[6]Darthmouth College, Department of Biological Sciences; Hanover, NH 03755, USA

[7]Department of Collective Behaviour, Max Planck Institute for Ornithology

[8]Chair of Biodiversity and Collective Behaviour, Department of Biology, University of Konstanz

[9]Edward Grey Institute of Ornithology, Department of Zoology, University of Oxford

*to whom correspondence should be addressed: PO Montiglio (makitsimple@gmail.com) and DR Farine (dfarine@orn.mpg.de)

All authors contributed equally

# Preface

Biological interactions span across levels of organisation. Because genes and phenotypes are nested within individuals, and individuals within populations, interactions within one level of biological organisation are inherently linked to interactions at others. Such nested networks highlight two central properties of biological systems: a) processes occurring across multiple levels of organization shape connections among biological units at any given level of organisation; b) ecological effects occurring at a given level of organisation can propagate up or down to additional levels. Explicitly considering nested network structure can generate new insights into key biological phenomena, and promote broader considerations of interactions in evolutionary theory.

# Abstract

Interactions are ubiquitous across biological systems. These interactions can be abstracted as patterns of connections among distinct units – such as genes, proteins, individual organisms, or species – which form a hierarchy of biological organisation. Connections in this hierarchy are arranged in a nested structure: gene and protein networks shape phenotypic traits and together constitute individuals, individuals are embedded within populations, populations within communities, and communities within ecosystems. This pervasive "nestedness" of networks can result in propagation of the effects from within-level interactions at one level to units at higher or lower levels of organization. The concept of nested biological networks is implicit in a variety of disciplines ranging from the study of genetic circuits regulating phenotypic trait expression (Babu *et al.* 2004), to the study of predator-prey interactions influencing community composition (Poisot *et al.* 2016). However, studies typically only address interactions within and among directly neighbouring hierarchical levels, such as genotypes and phenotypic traits, or populations and communities. Here, we formalise nested networks as having nodes that can contain, or be embedded in, other nodes, and where edges can bridge connections between sets of embedded nodes. We then argue that explicitly accounting for network nestedness across levels of organization will encourage integrative thinking on new interdisciplinary research fronts. We focus on two phenomena in particular: (i) indirect connections among units can arise from the structure of connections at higher or lower levels of organisation, (ii) the propagation of effects across neighbouring hierarchical levels of organization. This framework of nested interaction networks provides a tool for researchers across disciplines to conceptualize their work as elements on a common scaffold.

# Introduction

Within a level of biological organisation, units (e.g., genes, phenotypic traits, individuals, populations) affect each other's state and activity via various forms of interaction (e.g. protein-protein binding, pleiotropy, competition, mutualism), which we will refer to as connections between units. The overall patterns of connections among units shape gene circuits and phenotypic traits, as well as group-, population-, and community-level processes (Hendry 2017; Garcia-Callejas *et al.* 2018). Understanding the forms and consequences of such connections is a central focus of research disciplines spanning genetics (Boone *et al.* 2007; Mackay 2013; Civelek & Lusis 2014), development (Davidson & Erwin 2006), behaviour (Wilson *et al.* 2014), ecology (Poisot *et al.* 2015, 2016), and evolution (Moore *et al.* 1997; Agrawal *et al.* 2001; Bijma *et al.* 2007).

Recent studies have used various aspects of network theory to emphasize the key features of biological connections and their functional consequences. For example, representation of biological systems as multi-layer networks can highlight how units are connected through direct or indirect connections, or clarify relationships between units within and across different levels of organization (Ferrera *et al.* 2009; Kivelä *et al.* 2014; Pilosof *et al.* 2017). Research on food webs and eco-evolutionary dynamics, for instance, have highlighted how biological links connect contiguous levels of organisation (Basu *et al.* 2001; Knight 2007). Here, we emphasize that these concepts have much broader applicability: (i) networks not only span within but also across multiple levels of biological organisation, creating direct and – less intuitively - indirect connections between genes, phenotypic traits, organisms, etc., and (ii) the fact that units at one level of organization are nested within units at a higher-level results in propagation of network dynamics from one organization level to neighbouring levels and beyond.

A direct consequence of nested network structure is that connections among units at a given level of biological organisation are shaped by, or form the basis for, connections among units at higher and lower levels of organisation (Levin 1992). For example, trophic or competitive interactions among individual organisms can create connections between their phenotypic traits and genetic network structures (Foster *et al.* 2017). Similarly, emergent collective behaviour at organismal scales, shaped by the networks of connections among individuals (Rosenthal *et al.* 2015), fundamentally shape the patterns of connection at higher levels, such as the dynamics of food webs. As others have explored (Szathmáry & Smith 1995), hierarchical structure is a fundamental property of biological systems, and it is implicit on our understanding of how, for example, connections at the genetic, molecular and cellular level within a given organism lead to the emergence of phenotypic traits (Barrett *et al.* 2008; Bendesky *et al.* 2017), to how connections among organisms of different species structure ecological communities (Schmitz *et al.* 2008; Gounand *et al.* 2017). Analysing how nested networks operate can clarify short-term mechanistic questions – for instance, how gene-gene interactions can propagate through individual-individual engagements and alter large-scale ecosystem function. Analysing how these network structures change through time naturally leads to equally important questions about evolutionary dynamics on longer time scales.

In this paper, we discuss the utility of representing hierarchically embedded biological structures as nested networks (Figure 1). This approach is intended to recapitulate the natural hierarchy of levels of biological organization (Szathmáry & Smith 1995) from both mechanistic and evolutionary perspectives. We outline how nested networks are already present, either implicitly or explicitly, in research across sub-disciplines of biology, from microbiology to animal behaviour, and from molecular genetics to ecosystem research. However, we suggest that the nested network concept can be applied more formally, and we highlight two key important phenomena that it can capture. First, two or more biological units are often cryptically connected due to mutual interaction partners within or across levels of organization. Second, the innate nestedness of connections in biological networks inevitably causes effect propagation up or down through several levels of organization. We emphasize that this perspective can point to new insights into biological phenomena and encourage research transcending traditional disciplinary boundaries.

## Nested networks capture biological organisation across scales

Recent work in several disciplines within ecology and evolution has highlighted the value of network representations to understand the outcomes of biological relationships within and across layers of organization in a given system of study. In a generic network representation, biological entities are depicted using groups of units, or nodes (e.g., genes, phenotypic traits, individuals, populations, communities), and connections between nodes, represented as edges (Figure 1A, Table 1), to capture some biological relationship between units. Both nodes and edges can be described by a range of attributes (e.g., sex and age in the case of organismal units, and intensity or frequency in the case of predator-prey interactions). Nodes connected in a network typically represent units at the same level of biological organisation. The multi-layered network approach has recently extended this framework to account for multiple types of connections among units, or among multiple types of units (Pilosof *et al.* 2017). Further, the spatial and temporal dynamics of connections can be mapped in time-varying networks (Rosenthal *et al.* 2015). In molecular systems biology, transcriptional profile analysis, and metabolomics, network-based approaches have a rich history; knowledge about gene regulatory networks and metabolic networks has provided insights on genotype-phenotype relationships and how they evolve in response to selection (Stern *et al.* 2009; Olson-Manning *et al.* 2012). The combination of high-throughput sequencing techniques and network modelling has led to major improvements in prediction of microbial species interactions within microbiomes (Faust & Raes 2012). In behavioural ecology, social network approaches have helped quantify the likelihood and speed of transmission of social information (Aplin *et al.* 2015), or characterise the structure of dominance hierarchies with greater resolution and flexibility (Shizuka & McDonald 2012). At broader scales, studies of eco-evolutionary dynamics consider units spanning from genes to populations and the interconnections among them (Hendry 2017).

Importantly, networks at one level of organisation are inherently embedded into nodes at higher levels of organisation (Figure 1B). For example, genotypes and the protein networks they encode underlie phenotypic traits, which altogether constitute individual organisms that themselves live and interact together in populations, whose relationship to other populations defines ecological communities. This inherent nestedness means that connections at lower levels can facilitate or contribute to connections at higher levels of organization (see Figure 1C). To take a familiar example, the dispersal of individuals creates connections between populations in the form of gene flow. Connections at higher levels of organization, on the other hand, inevitably lead to connections among their units at lower levels of organization. In a further example, social interactions between two individuals may lead to indirect connections between each individual's network of physiological states, structural phenotypic traits, and internal networks of gene transcription and translation. Recent approaches have conceived of populations and communities as multi-layered networks of networks (Kivelä *et al.* 2014; Pilosof *et al.* 2017). These allow more than one type of interaction network to be overlaid and integrated on the same set of nodes, or the nodes of two or more networks to be connected to each other. The concept we advance here adds to these insights by emphasizing that we can decompose connection patterns at a given level of organisation as the result of connections and node properties of units at lower and higher levels of organisation.

The details and nomenclature of a network representation will depend on the study system in question, but in general, connections between any two network nodes can be physical and direct (e.g. a mating event between two individuals) or more abstracted and indirect (e.g. exploitation competition between two social groups). Within each level of biological organisation, multiple categories of nodes and types of connections can exist (Figure 1D, see also Pilosof *et al.* 2017). For example, a community unit can contain units representing individuals of a prey species (Category A) and individuals of a predator species (Category B). Units such as individual organisms are typically connected to others within or across different categories, and this may generate unexpected connections among populations that ordinarily would not be considered to interact with each other. Helping to identify counter-intuitive links between individuals, populations, or communities that at first glance appear unconnected is arguably the central benefit of a nested network perspective. For example, predator populations that forage across spatially or temporally disconnected prey populations in a food web can generate an indirect link between the prey populations (Bjørnstad *et al.* 1999; Ims & Andreassen 2000) (Box 1). These inter-population indirect connections then allow for the behaviours, structural phenotypic traits, and underlying genetic/development networks of individuals across the prey populations to influence one another on ecological and evolutionary time scales (per Figure 1A). Such a representation highlights that changes in the behaviour of one prey population can propagate across levels by affecting the local predation pressure and, therefore, the pattern of natural selection that the other prey population experiences. In Box 2, we present a simple model to illustrate how the indirect connections created among prey populations by predators can impact the evolution of prey phenotype. The model also shows how changes in phenotypes within prey populations shape the network of interactions among predators.

The interaction structure of a population is determined both by the phenotypic traits of its individual members and the population's position within the broader ecological community. Thus, many phenotypic traits expressed within populations have cascading effects on community dynamics (Bolnick *et al.* 2011). For example, Trinidadian guppies from populations that coexist with or without dangerous predators differ in their growth, fecundity, and resistance against parasitism (Stephenson *et al.* 2015). Connections between communities create connections between populations, and eventually individuals, their phenotypic traits, and their genes. Model systems in evolutionary biology, such as the three-spine stickleback, have been used to investigate a large variety of genetic, behavioural, population-focused, and community level phenomena that can be integrated through a nested network perspective (Figure 2). Another common example of linkages across levels of organization is the dependence of disease transmission on the network structure of hosts, where the connections among hosts depend on the genes (Radersma *et al.* 2017), phenotypic traits (Mason 2016), and inter-individual relationships (VanderWaal *et al.* 2014), and the transmission dynamic plays out at the scale of populations and communities (Penczykowski *et al.* 2016).

## Changing networks across generations

Accounting for the natural nestedness of biological networks, and how and why they can change in time, also promises to deepen our understanding of evolution in natural environments. Within each level of a nested biological network, changes in the environment can alter (i) the attributes of each unit, (ii) the attributes of connections, (iii) the pattern (or strength) of connections among units, and/or (iv) the number of units. For example, environmental shifts can couple or decouple phenotypic traits (such as growth and survival, van Noordwijk & de Jong 1986); alter the size of groups that the environment can support (Lima *et al.* 2002); alter the kinds of interactions that occur among populations (e.g., a shift from no interaction to predator-prey interaction when a predator's other food sources become scarce, Lima *et al.* 2002); or fundamentally shape meta-community structure (Holt *et al.* 2003; Leibold *et al.* 2004).

To develop a well-known example using this framework, consider a host-parasite system in which two interacting networks exist at the organismal level: the network of hosts, and the network of parasites, both of which are closely engaged with each other. In addition to these community and population level network interactions, each host and parasite contains within it several more nested networks of phenotypic, developmental, and genotypic interactions. Evolutionary change in a host characteristic (e.g., an increase in the constitutive investment into an immune response) can alter the connections among and within all the component networks in the system. Host evolutionary change can exert selection on the parasites, which may evolve in response and in turn exert selection on hosts with respect to the phenotypic traits influencing interactions with the parasite. This evolutionary arms race (Decaestecker *et al.* 2007) can influence host phenotypic traits that then feed into host-host interaction networks – for example, parasitism affects male nuptial colouration in three-spined sticklebacks, and the resulting variation in nuptial coloration among males provides the substrate for sexual selection exerted by females' preference for increased male colouration (Milinski & Bakker 1990). Furthermore, parasitic interactions within a host can influence both parasite-parasite and host-parasite interactions, such as when ectoparasites alter host health and lower barriers to secondary bacterial infections (Bandilla *et al.* 2006; Pedersen & Fenton 2007). In the other direction, changes in host characteristics including parasite tolerance can also affect the abundance of both host and parasite populations (Hassell & Waage 1984; Palkovacs & Post 2009; Penczykowski *et al.* 2011, 2016), and the connections among the populations in their community (Wood *et al.* 2007; Lafferty *et al.* 2008). Evolutionary changes in host-parasite interactions can also affect other species in the community, both directly and indirectly. For example, increased average parasite load in a predator population decreases host viability and reduces predation pressure on a corresponding prey population, thus affecting that prey population's interactions with its own food sources.

This example highlights how changes at a given level of organisation can exert influence within and across all levels of organization in the system. However, to date studies have primarily focused on the effects of change within one such level of organization, or, at most, two adjacent levels. Explicitly thinking about nested networks, where dynamics in a given study system are linked to flow-on changes across the broader environment in which that system is embedded, will encourage a broader and more interdisciplinary general approach to behavioural and community ecology.

## Future Work

Though the concept of nested networks has been touched upon implicitly or explicitly by a number of sub-disciplines in ecology and evolution, accounting for the nested structure of biological systems in the theory and practice of behavioural ecology and community ecology remains rare. An immediate obstacle is to devise new quantitative approaches to measuring network propagation and indirect connection effects in both theoretical and experimental frameworks. Models could provide testable predictions about how we might expect connections at one scale to influence connections at higher levels, and vice versa (Moore *et al.* 1997; Wolf *et al.* 1999; McGlothlin *et al.* 2010). While the advantages of interdisciplinary science are clear, one potential challenge is to ensure that all researchers can find a common ground. This is no trivial task, and the conceptual framework discussed here is in part meant to provide a stepping-stone for quantitative integration of nominally different concepts of biological interactions.

Finding suitable model systems for studying nested network dynamics is an important future goal, as well. The ideal system should amenable to experimentation across all levels of organization from genetics, to individuals, to populations and communities. We have highlighted several examples that demonstrate that this is possible. Microbial systems should be especially powerful given the experimental control they provide. Host-parasite systems also offer an obvious link between connections among genes and connections among individuals and across these levels (Milinski & Bakker 1990; Susi *et al.* 2015; Penczykowski *et al.* 2016). Finally, the work done on sticklebacks (Box 1) and other teleost fish (e.g. cichlids and guppies) exemplifies how our framework can integrate top-down approaches in natural environments with bottom-up approaches using genomic tools available for these species.

## Conclusions

A wealth of cutting edge research informs our understanding of molecular mechanisms, ecological dynamics, and evolutionary processes at different levels of biological organisation. We argue that explicitly considering the nestedness of biological networks will resolve new interdisciplinary questions on how more mechanism-inclined versus ecology and evolution-inclined research programs relate to each other. Among the most pressing obstacles to progress is to bring together multiple fields analysing the structure of connections at each level of biological organisation and from multiple perspectives, which normally operate under different vocabularies and focus on different spatial and temporal scales. We aimed here to develop a step toward tackling this obstacle by providing a conceptual tool that can align different traditions of analysis and enquiry to generate truly novel insights.

**Box 1: A case study of nested networks of sticklebacks and their predators**

Three spine stickleback *Gasterosteus aculeatus* are found in aquatic environments across the northern hemisphere. Populations originated in the marine environment and have repeatedly colonized freshwater habitats since the last glacial retreat. The interactions between these populations and a range of predators have shaped numerous aspects of their morphology, behaviour, and life histories (Reimchen 1994; Walker 1997; McKinnon & Rundle 2002; Bell 2005), making them a classic model system in evolutionary biology. We can conceptualise stickleback populations as two broad communities: a marine and a freshwater community. These communities are connected by anadromous stickleback populations that live in marine and reproduce in freshwater environments. The two communities differ in their primary predators: marine communities are dominated by large piscivorous predators, whereas larval odonates (dragonflies and damselflies) are predominant predators in many freshwater communities (Reimchen 1994).

Stickleback populations within each community differ in several traits, an important one being their defensive spines and armour plating. In marine sticklebacks, morphological defences are beneficial against piscivorous predators (Bell *et al.* 1993; Marchinko 2009). However, these are costly to freshwater sticklebacks because odonates can more easily grasp the spines and plates, resulting in the repeated loss of these traits in freshwater populations (Bell *et al.* 1993). Two genes, Pitx1 and Eda, explain much of the variation found in armour plating, pelvic girdle, and dorsal spines (Shapiro *et al.* 2004; Colosimo *et al.* 2005; Barrett *et al.* 2008; Marchinko 2009; Chan *et al.* 2010). These genes form a gene interaction network within an individual (Peichel 2005; Jones *et al.* 2012), and thus are nested within phenotypic traits that are contained within the individuals. Individuals interact and form populations within communities.

Understanding the connections among communities and populations is key to gaining an accurate picture of underlying drivers of evolutionary processes. For example, overfishing of mesopredators in marine environments is a global problem (Botsford *et al.* 1997; Pauly & Palomares 2005). Reductions in predator populations could lead to an increase in the adult survival of anadromous sticklebacks in the marine community. This increase could result in more individuals returning to the freshwater community, and a higher frequency of alleles that generate defensive morphology.

Dragonfly populations are then likely to benefit. If adult dragonflies choose to remain at the location where this armoured population is found, this could decrease predation on nearby freshwater stickleback populations. Alternatively, if an increase in the dragonfly population creates a source of dispersing adult dragonflies, this could result in increased predation in other freshwater populations, thus driving stronger selection against alleles encoding defensive spines and plates. In either case, an indirect link can exist between otherwise disparate marine and freshwater stickleback populations, as well as between different freshwater stickleback populations. This example demonstrates how exploring nested networks can facilitate discussions among disciplines (in this case conservation specialists, population ecologists, and evolutionary biologists), and enable the development of novel hypotheses and testable predictions.

## Box 2. Nested networks explain variation in phenotypic correlations among populations

Accounting for the nestedness of interactions among biological units can provide novel insights into key eco-evolutionary questions. For example, predation often leads to the evolution of correlated suites of traits in prey populations. Exposure to predators leads to correlations between body size and diel migration patterns in Daphnia (Demeester *et al.* 1995) and size at maturity and amount of color in guppies (Endler 1995). However, antipredator traits are correlated in some populations and not in others and it is often unclear when and why. A key insight though is that the pattern of selection exerted by predators will be at least partly determined by interactions at higher levels of biological organization, such as among the predator populations themselves. For example, physically isolated vole populations exhibit population synchrony due to predation pressure from shared avian predators (Ims & Steen 1990). By explicitly accounting for the structure of connections within and among predator and prey populations, the nested network framework offers a better understanding of the processes driving patterns of predator selection on correlated traits in prey populations. Here we construct a toy model to illustrate how this framework allows us to make predictions about how connections at one level, e.g. connections among predator populations, drive patterns of connections at other levels, e.g. evolution of correlated traits within prey populations.

We start with six physically isolated prey populations, for example ponds with fish. We assume no migration among populations of prey. Mobile predator populations (e.g. birds) hunt on these ponds. The home range of each predator population overlaps several ponds, and predator populations distribute their foraging effort according to an ideal free distribution. Thus, populations of predators create indirect connections among prey populations (i.e. a network of ponds; Fig 1a). We model the spread of two independent loci in a single prey pond within four different predator network structures (see supplementary materials for more details). Locus 1 encodes a trait where the dominant allele A encodes resistance to predation (versus vulnerability for allele a) but at a fitness cost. Locus 2 encodes a trait where the dominant allele B provides a fitness benefit but makes individuals conspicuous to predators (versus cryptic for allele b); hence we assume a rare morph disadvantage for this trait. Predators allocate their foraging effort across ponds in response to (i) the frequency of the resistant prey and (ii) the intensity of foraging by other predators, in the ponds in which they forage.

Allowing connections among predator populations alters the evolution of prey (see supplementary material). The position of the pond within the network of connections created by predation determines which prey phenotype evolves to become the most frequent and the correlation between resistance and conspicuousness (Fig 3). When the alleles arise in a pond at the periphery, predation maintains low frequencies of resistant and conspicuous prey while favoring vulnerable and conspicuous individuals. In contrast, in a pond adjacent to the periphery, predation favors resistant and cryptic individuals and, to a lower extent, resistant and conspicuous individuals. Finally, in a centrally positioned pond, predation leads to the fixation of resistant and cryptic individuals only. Thus, our model shows that even though all prey ponds experience predation, we should not expect predation to drive the same patterns of phenotypic correlations in each pond. Rather, the patterns of correlation among traits in a given prey population are determined by how the population is embedded within the larger metapopulation network. Many models have analyzed prey evolution in response to predation. This simple model shows that adopting a nested network framework can encourage us to consider connections across more than one level of organization to improve our ability to explain why predation seems to have such a wide range of consequences for prey.

Interestingly, changes in allele frequencies within ponds of prey propagate to higher levels of organization, shaping the distribution of predation pressure across ponds, and thus the network of competitive interactions among predator populations (Fig. 4). The exact impact of changes in allele frequencies vary according to the position of the pond in which the alleles arise. Increasing the frequencies of either allele A or allele B generally makes a pond less profitable. As a result, predators avoid such ponds and concentrate their efforts on the remaining ponds. If alleles arise in a peripheral pond (e.g. Fig. 4, left panel), then the predators mostly re-distribute their effort evenly and equally among remaining ponds. However, if alleles arise in a more central pond, then the predator networks that arise are much more unpredictable. In some cases (e.g. Fig. 4, middle panel) allele B arising in a pond increases the competitive interactions among predators by generating stronger predation pressure in some ponds and little in others (i.e. they become specialist). If alleles arise in the most central pond (e.g. Fig. 4, right panel), the same increase in allele frequency leads to predators foraging more evenly across all ponds. As a result, the network of competitive interactions includes fewer interactions among predators with lower intensity; that is all ponds experience more similar predation pressure.

# Table

Table 1: Types of connections within nested networks

Connections between units can represent a range of direct and indirect relationships. Often, these connections represent the outcome of direct physical contact between two units, such as conjugation between two bacteria. Alternatively, two units can be indirectly connected, for example, if they are influenced by separate interactions with the same third party (Figure 1). Finally, these connections can be relatively temporary like a mating event between two animals or more stable, such as when two proteins generate a stable polymer. Here we list a few examples of potential direct and indirect connections between units at different broad levels of biological organisation.

| Level of biological organisation | Example unit categories | Example connections |
| --- | --- | --- |
| Subcellular | Genes | The expression of one gene produces a transcription factor that alters the expression of other genes. |
| | Protein complexes | Many proteins assemble into multi-component structures, such as the flagellar basal body |
| Phenotype | Behaviours | The expression of parental care behaviour is connected to the expression of aggression due to the levels of particular hormones. |
| | Morphology | The length and shape of one limb is connected to the length and shape of the other limb through genetic and/or physiological mechanisms. |
| Individual | Bacterium | One bacterium secretes a substance that has a detrimental (or beneficial) effect on another. |
| | Individual animals | Two female zebra finches are connected by both having mated with the same male zebra finch |
| Population | Populations of the same species | Two physically isolated populations of crabs are connected to each other via predation by the same |

|  |  | population of gulls |
|  | Populations of different species | A population of bacteria is connected to a second population of bacteria because it produces a substance that augments the growth of the second population. |
| Ecosystems | Community | A terrestrial community is connected to an aquatic community via nutrient cycling processes. |

# Figures

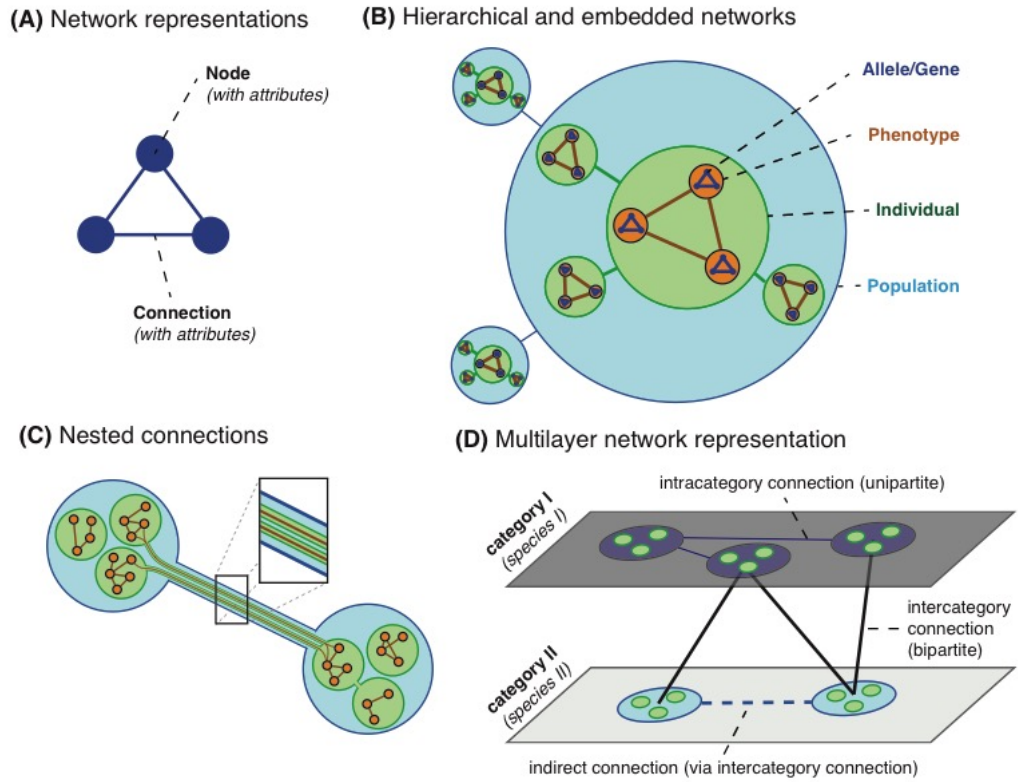

Figure 1: From genes to communities: a framework for describing nested networks of connections.

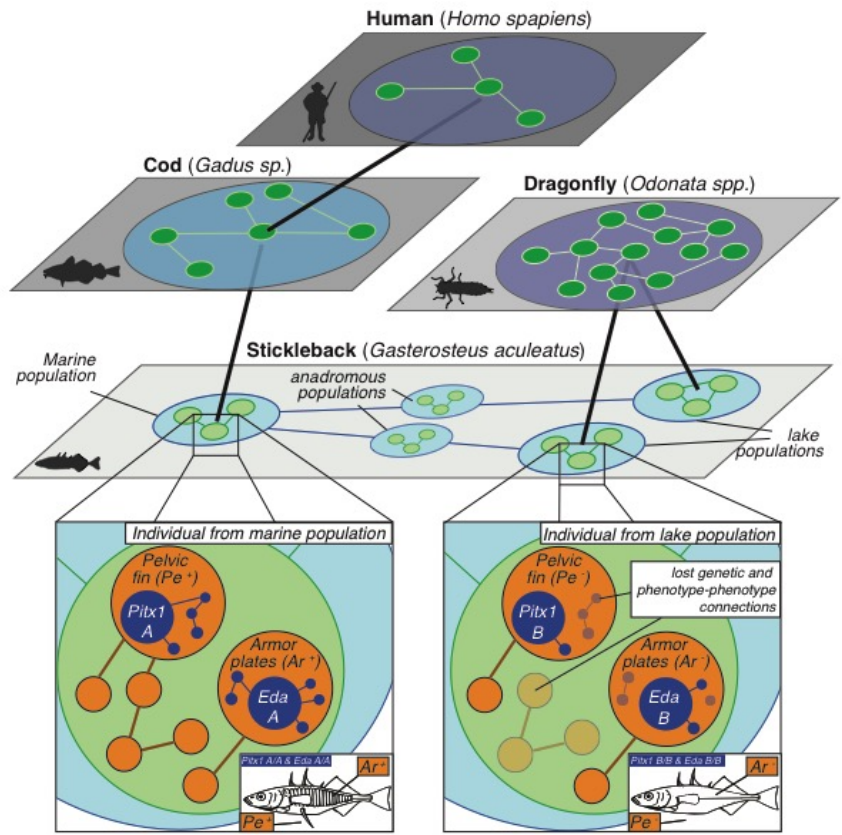

Figure 2: The stickleback system as a case study for nested networks and interactions across hierarchical and species levels.

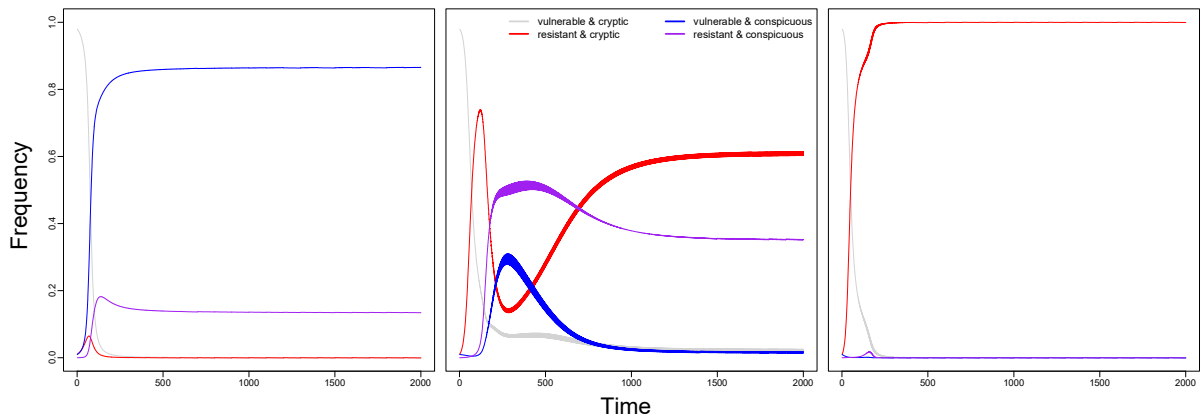

**Fig 3.** Changes in prey phenotype frequencies over time, within a pond at the periphery (left panel), near the periphery (middle panel) or centrally positioned (right panel) as a result of predation. The lines indicating the frequencies become thicker when the frequencies cycle.

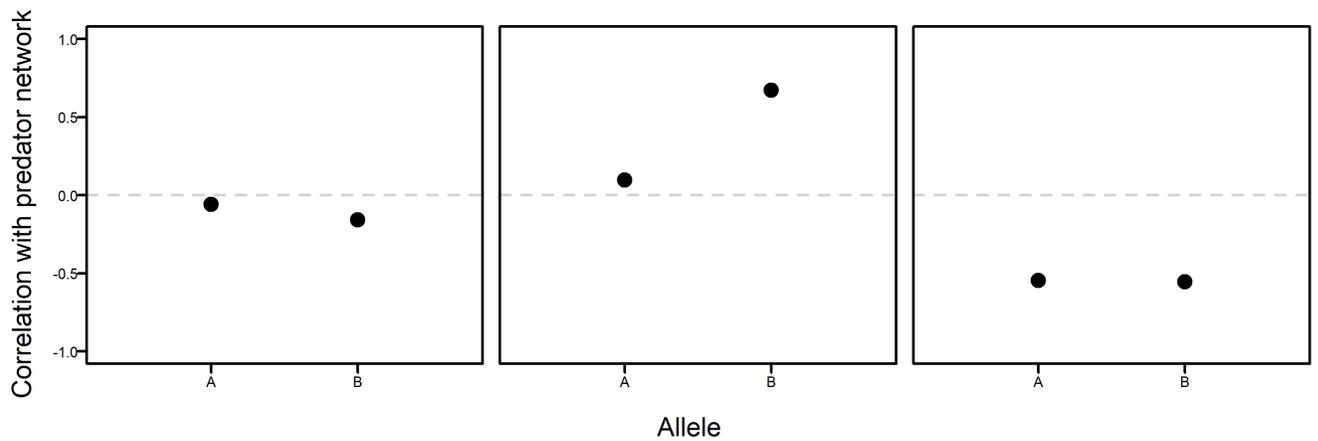

**Fig. 4.** Correlation between the frequency of alleles A and B in the ponds of prey and the coefficient of variation (CV) of edge weights from the network of competitive interactions among predators. A positive correlation between allele frequency and coefficient of variation for edge weight (e.g. middle panel) means that increases in allele frequency forces predators to spread their foraging effort unequally among the ponds that they exploit. A negative correlation between allele frequency and coefficient of variation for predation effort (e.g. right panel) means that increases in allele frequency evens out how predators exert their pressure across ponds. Correlations are calculated using the Pearson's correlation coefficient of each allele frequency and the relative CV value for each step in the first 500 time steps of the model (when most systems are changing).

# Supplementary Materiel 1. The Model

**Constant predation**

We model a series of isolated diploid prey populations, each of which is infinite in size. Prey are polymorphic at two loci. Locus 1 controls whether individuals express a character that makes them resistant to predation (allele A, dominant) compared to wild type vulnerable individuals (allele a, recessive). Locus 2 controls whether individuals express a character that makes them more conspicuous (allele B, dominant) compared with wild type cryptic individuals (allele b, recessive). Thus, the prey population can consist of four different phenotypes: resistant and conspicuous (genotypes AABB, AaBb, AABb, AaBB); resistant and cryptic (AAbb, Aabb); vulnerable and conspicuous (aaBb, aaBB), and vulnerable and cryptic (aabb). We assume these genotypes mate randomly and that no mutation occurs. Because populations are infinite in size, we assume no genetic drift.

Each pond is preyed upon by a population of predators. Individuals possessing the allele A accrue fitness benefits through lower predation rates, but they also pay a fixed fitness cost ($c$). Such a cost could arise, for example, because producing the antipredatory character is associated with reduced fecundity or a slower growth rate. Thus, the fitness effects of allele A are $\alpha_i - c$, where $\alpha_i$ is the predation pressure exerted on pond $i$. Individuals bearing allele B accrue fitness benefits ($d$) in another context (e.g. increased fecundity) but they also pay a fitness cost through the oddity effect: predators preferentially prey upon these individuals when they are rare because they are conspicuous and stand out among individuals in their pond. The oddity effect is proportional to the intensity of predation pressure. The fitness effects of allele B are $d - \alpha_i e^{-10p(B)}$, where $p(B)$ is the proportion of conspicuous individuals.

We assume that the two loci are independent in their fitness effects. Thus, the relative fitness of the genotypes (relative to that of the wild type genotype aabb) is as follows:

$$W_{AABB} = W_{AaBB} = W_{AABb} = W_{AaBb} = 1 + \alpha_i - c + d - \alpha_i e^{-10p(cB)}$$

$$W_{AAbb} = W_{Aabb} = 1 + \alpha_i - c$$

$$W_{aaBB} = W_{aaBb} = 1 + d - \alpha_i e^{-10p(B)}$$

$$W_{aabb} = 1$$

We adopt a population genetics approach where each generation starts with the proportion of each gamete (at generation $t$), which assemble randomly to generate the frequencies of adult genotypes ($t$ prior to selection, $t_{-s}$ hereafter). The proportion of each genotype is then adjusted as a function of its fitness W ($t_s$), before generating the gametes starting the next generation ($t + 1$). For example, the frequency of genotype AABB is the product of the frequency of this genotype at the previous generation and of its fitness relative to that of the population.

$$p(AABB)_{t_s} = p(AABB)_{t_{-s}} \frac{W_{AABB}}{p(AABB)_{t_{-s}} W_{AABB} + p(AaBB)_{t_{-s}} W_{AaBB} + \cdots + p(aabb)_{t_{-s}} W_{aabb}}$$

The new proportion of each genotype is then used to compute the frequency of each gamete type (AB, aB, Ab, and ab) seeding the next generation.

Under constant and sufficient predation pressure ($\alpha_i$), being resistant incurs higher fitness (i.e. when $\alpha_i > c$), and thus, the resistant trait (Allele A) goes to fixation. Similarly, under constant and sufficient predation when $\alpha_i e^{-10p(B)} > d$, conspicuous individuals are extirpated from the population and allele b is pushed to fixation.

**Accounting for interactions among predators**

Now suppose that the predator populations can forage on more than one pond and that multiple predator populations can forage on the same pond. This network of predator populations preying upon different ponds can vary in our model. Suppose also that the dominant alleles at each locus (A and B) can appear in only one of these ponds (all the other ponds consist of only wildtype aabb individual prey). We assume that all predator populations have equal effects on the prey (i.e. a predator from one population is as effective as a predator from a different population), and that they compete against one-another. Thus, we model that each predator population adjust its investment in each pond they have access to as a function of the investment of the other predator populations preying upon the same pond. In other words, predator populations forage across the ponds they have access to following an ideal free distribution.

We assume that predators also adjust their foraging effort on a pond as a function of the frequency of resistant individuals. As the frequency of resistant prey increases in the focal pond (in response to predation pressure), foraging becomes less and less profitable for the predator populations who prey upon that pond. These predator populations therefore reduce their investment in that pond and redirect their efforts to another pond they have access to. Thus, for pond $i$ that contains the evolving prey, the investment for a given predator $j$ at generation $t+1$ in pond $i$ will be:

$\alpha_{i,j,t+1} = \frac{1-p(A)}{\sum_{j=1}^{N} \alpha_{i,j,t}}$, where $N$ is the number of predator populations that target that pond. The denominator of this formula expresses the idea that, as the foraging effort of other predator populations on that pond increases, a focal predator population will reduce its foraging effort in that pond (and vice versa). Shifting investment to and from ponds increases and decreases predation pressure in other ponds. The numerator of this formula expresses the idea that as the frequency of resistant prey increases, the frequency of profitable prey decreases in the pond. Predators should thus lower their foraging effort on a given pond as the frequency of resistant prey increases. Each predator effort is then rescaled to always sum to 1, thus maintaining constant overall predation across the entire system at all times (i.e. predator populations are assumed infinite and do not fluctuate in size). This allows us to compare the effects of changing the structure of connections among ponds without the confounding effects of increasing or decreasing overall predation pressure on the network of ponds.

The result of the above allocation and re-allocation of foraging effort by predator populations is that the predation pressure on a specific pond can fluctuate over time due to asynchrony in the foraging effort among predator populations (e.g. if $\alpha_{i,j,t}$ is low, predators will increase their foraging effort at the next time step, resulting in $\alpha_{i,j,t+1}$ becoming high, in which case they again reduce effort in $\alpha_{i,j,t+2}$, and so on). Fluctuations in predation pressure on a pond should also lead to changes in the frequency of the dominant alleles for each locus. When prey populations are released from predation pressure on the focal pond, the frequency of resistant prey (allele A) should go back down again because the cost ($c$) will exceed the benefits ($\alpha_i$). Likewise, lower predation pressures in that pond should also increase the frequency of conspicuous individuals (allele B) because the fitness costs ($\alpha_i e^{-10p(B)}$) will be lower than the benefits ($d$).

These cycles of fluctuations in the frequency of resistant and conspicuous prey can further enhance cycling in the foraging effort by predator populations. Similarly, in some of the ponds, fluctuations can lead to fixation of the frequency of conspicuous prey, whereas in others, such conspicuous prey are completely extirpated. Whether predation leads to fixation of resistance and conspicuousness in the pond, to their extirpation from the pond, or to their maintenance at intermediate frequencies, and whether the two traits are linked will therefore depend on the number of ponds, the number of predator populations, and on the amount of overlap among predators.

The following figures show the effect of such indirect connections among ponds on the evolution of frequencies in alleles A and B across a range of values for $c$ and $b$, and fluctuations in predator investment $\alpha_i$ over generations.

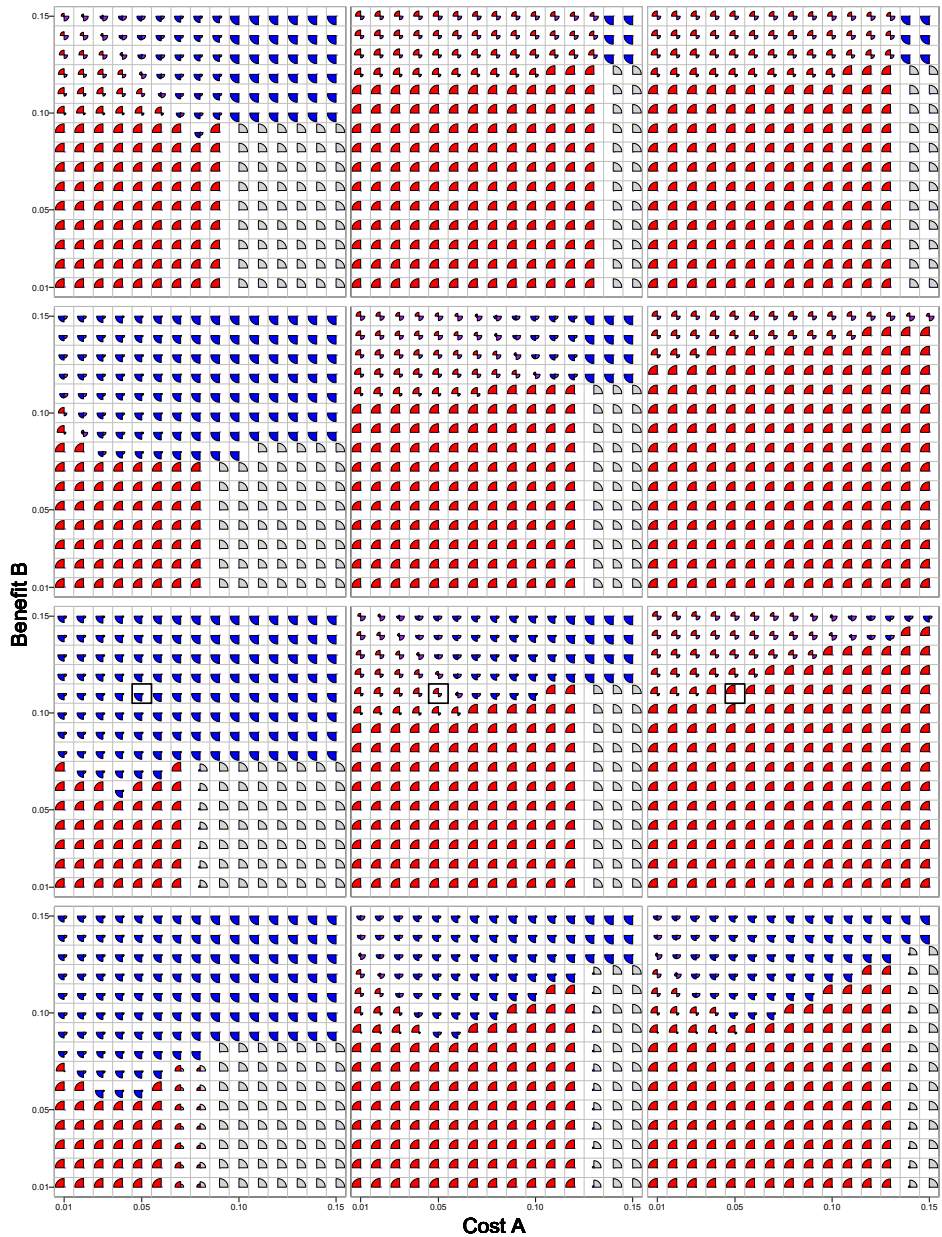

**Fig. S1.** Example of the emergence of different population genetic structures under different costs for allele A and benefits for allele B. All simulations were run in an environment containing 6 ponds, with the lower row representing a system where 5 predator populations are each connected

to 2 ponds, the second row represents a system where 4 predators are each connected to 3 ponds, and so on (i.e. predator $degree = 2, \ldots, 5$ from bottom to top). In each case, the efficacy factor was increased to account for the reduction in the number of predators (thus maintaining the overall predator pressure the identical in all simulations). Each column represents a given focal pond, with the left-most column representing the most peripheral pond (connected to $N = degree - 1$ predators), the middle column represents the second-from-left pond (connected $N = degree$ predators), and the right-most column represents the third-from-left pond (also connected $N = degree$ predators). The wedges on the plot represent the fraction of the population having that phenotype, with the colours matching those from Box 3 in the main text (grey=vulnerable & cryptic, red=resistant & cryptic, blue=vulnerable & conspicuous, purple=resistant & conspicuous). The size of the wedges in each 'pie' sum to 1.

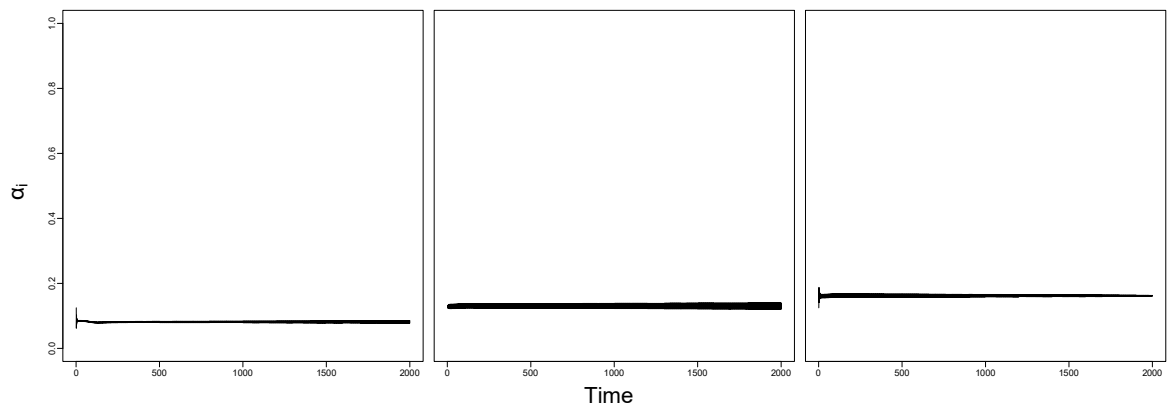

**Fig. S2.** Predator pressure in ponds 1 to 3 when the alleles are present in pond 1. That is, plots show the values of $\alpha_i$ for the left-most plot in the figure given in Box 3. The thickness of the line represents the breadth of the cycling of the value of $\alpha_i$ from one generation to the next.

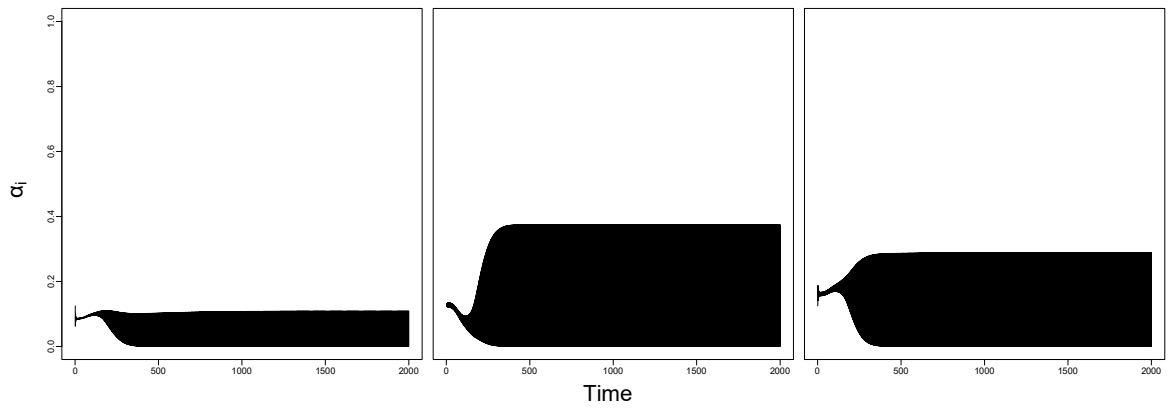

**Fig. S3.** Predator pressure in ponds 1 to 3 when the alleles are present in pond 2. That is, plots show the values of $\alpha_i$ for the central plot in the figure given in Box 3. The thickness of the line represents the breadth of the cycling of the value of $\alpha_i$ from one generation to the next.

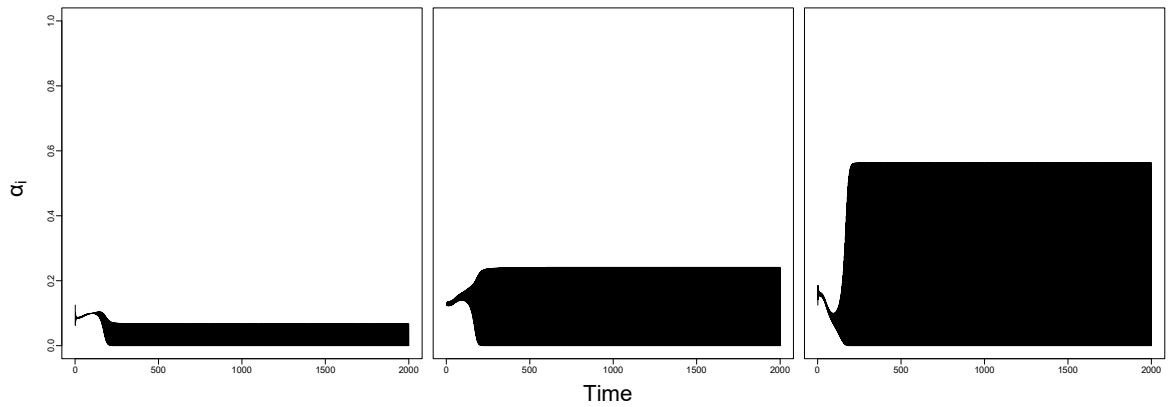

**Fig. S4.** Predator pressure in ponds 1 to 3 when the alleles are present in pond 3. That is, plots show the values of $\alpha_i$ for the left-most plot in the figure given in Box 3. The thickness of the line represents the breadth of the cycling of the value of $\alpha_i$ from one generation to the next.